\documentclass[fleqn]{llncs}

\usepackage{cite}
\usepackage{caption}

\usepackage{amsmath} 
\usepackage{amssymb}
\usepackage{wasysym}
\usepackage{mathtools}
\usepackage{mathrsfs} 

\usepackage{verbatim}

\newcommand{\lola}{Lola}
\newcommand\donotshow[1]{}
\usepackage{listings}

\usepackage{adjustbox}

\usepackage{amsthm}
\newtheorem{defs}{Definition}

\usepackage{listing}
\usepackage{placeins}
\usepackage{url}

\usepackage{tikz}
\usetikzlibrary{matrix}
\usepackage{wrapfig}

\usepackage[finalizecache=false, frozencache=true]{minted}

\begin{document}

\title{Stream Runtime Monitoring on UAS\thanks{Partially supported by the European Research Council (ERC) Grant OSARES (No.\ 683300) and by the German Research Foundation (DFG) as part of the Collaborative Research Center ``Methods and Tools for Understanding and Controlling Privacy'' (SFB 1223).}}

\author{Florian-Michael Adolf\inst{2}\and Peter Faymonville\inst{1} \and Bernd Finkbeiner\inst{1} \and Sebastian Schirmer\inst{2} \and Christoph Torens\inst{2}}

\institute{Saarland University, Reactive Systems Group \and DLR (German Aerospace Center), Institute of Flight Systems}

\maketitle

\begin{abstract}
  Unmanned Aircraft Systems (UAS) with autonomous de\-cision-making capabilities are of increasing interest for a wide area of applications such as logistics and disaster recovery. In order to ensure the correct behavior of the system and to recognize hazardous situations or system faults, we applied stream runtime monitoring techniques within the DLR ARTIS (Autonomous Research Testbed for Intelligent System) family of unmanned aircraft. We present our experience from specification elicitation, instrumentation, offline log-file analysis, and online monitoring on the flight computer on a test rig. The debugging and health management support through stream runtime monitoring techniques have proven highly beneficial for system design and development. At the same time, the project has identified
 usability improvements to the specification language, and has influenced the design of the language.
\end{abstract}

\section{Introduction}

Aerospace is an internationally harmonized, heavily regulated
safety-critical domain.  Aircraft development is guided by an uncompromising
demand for safety, and the integration of unmanned aircraft systems
(UAS) into populated airspace is raising a lot 
of concerns. As the name suggests, there is no human pilot on board an
unmanned aircraft.  The pilot is replaced by a number of highly automated systems,
which also ensure proper
synchronization with a base station on the ground. The
correctness and stability of these systems is critical for the safety
of the aircraft and its operating environment. As a result, substantial efforts in verification and
validation activities are required to comply with standards and the
high demand for functional correctness and safety.

Runtime verification has the potential to play a major role in the
development, testing, and operational control of unmanned aircraft.
During development, debugging is the key activity.  Traditionally,
log-files are inspected manually to find unexpected system behaviors.
However, the manual analysis quickly becomes infeasible if multiple
interacting subsystems need to be considered or if complex
computations have to be carried out to correlate the data.  Runtime
verification can automate this task and thus make debugging
dramatically more efficient.
During testing, runtime verification can be used to monitor
functional correctness of the system behavior.  Runtime
verification can be integrated into software- and hardware-in-the-loop
simulations as well as into full-scale flight tests. In contrast to simple
unit testing, a property that is formalized into a runtime monitor can
thus not only be tested in a test fixture, but reused in all test phases.
During operation, runtime verification can be used to monitor the
system health and the validity of environment assumptions. The reason
that system failures happen during a flight is often not because of
implementation errors, but because unforeseen events occur and the
requirement assumptions are no longer valid.  Integrating runtime
monitoring into operational control makes it possible to enforce
safety limitations, such as constraints on the altitude, speed,
geographic location and other operational aspects that increase the
risk emerging from the unmanned aircraft.  If all else fails, the
runtime monitor can also initiate contingency procedures and
failsafes.

In this paper, we report on a case study, carried out over
approximately one year in a collaboration between Saarland University
and the German Aerospace Center (DLR).  The goal has been to integrate
runtime monitoring into the ARTIS (Autonomous Research
Testbed for Intelligent Systems) platform. ARTIS consists of a versatile software
framework for the development and testing of intelligent functions, as well as a fleet of unmanned aircraft, which include several classes and sizes
of aircraft with different technical equipment and resulting
autonomous capabilities \cite{adolf07infotech,Torens2014}. 

Integrating runtime monitoring into a complex flight operation
framework like ARTIS is a profound challenge. An immediate observation
is that the data to be monitored is complex, and typically requires
nontrivial processing, which necessitates a highly expressive
specification language.  At the same time, there is no separation on
the hardware level between the flight control operations and the
monitoring engine. Performance guarantees for the monitoring code, in
particular with respect to memory usage, are therefore critically
important. Finally, we observed that there is a highly beneficial feedback
loop between the results of the monitoring and the continued design of
the aircraft. The specifications used for monitoring are increasingly
used as a documentation of the expected environment conditions and the
legal system behavior. Clear, modular specifications that can
easily be reused and adapted are therefore a key requirement.

Our runtime verification approach is based on the Lola specification
language and monitoring engine~\cite{Angelo+others/05/Lola,DBLP:conf/rv/FaymonvilleFST16}.
Lola specifications translate input streams, which contain sensor information
and other real-time data, into output streams, which contain the processed
sensor information and statistical aggregates over time. While Lola is a
very expressive specification language, it also comes with strong
performance guarantees: the efficiently monitorable fragment, which essentially consists of the
full language except that
unbounded lookahead into the future is not allowed, can be monitored
with constant memory. In principle, Lola is thus clearly in a good position for
the task at hand. If Lola would be sufficiently expressive for the monitoring of unmanned aircraft,
and whether Lola specifications would be sufficiently modular and understandable
for the interaction with the developers, seemed far from obvious at the outset
of our study.

The results reported in this paper are very positive. While Lola in
its as-is state was in fact not sufficiently expressive, the
integration of the missing features, essentially floating point arithmetic and trigonometric
functions, turned out to be straightforward. Lola's organizational
principle, where more complex output streams are computed in terms of
simpler output streams, was adapted easily by developers, who
appreciated the similarity to synchronous programming. Small and, in
retrospect, almost obvious additions to the language, such as a \textsf{switch}
statement for the description of state machines, made the
specification of the properties of interest significantly more natural and
elegant.

The shortage of published case studies is an often lamented problem
for research in runtime verification. We hope that our 
formalization of common properties of interest in the monitoring of
unmanned aircraft can serve as a reference for formalizations in other
runtime verification frameworks.  The major lesson learned in our work
is that while the development of such specifications is extremely
difficult and expensive, the benefits in terms of a more effective
debugging and testing process, and a mathematically rigorous
documentation of the expected system behavior are immense. Conversely,
from the perspective of the developer of a runtime verification tool,
the insights gained into the practical relevance of various language
features are similarly invaluable.

\donotshow{
Stream-based specification languages, such as Lola \cite{Angelo+others/05/Lola,DBLP:conf/rv/FaymonvilleFST16} are a modular and compositional specification mechanism for runtime monitoring. Their equation-based specification style and rich type- and function support as well as their support for statistics together with an efficient evaluation mechanism make them an ideal candidate for use in practice. 

Unmanned Aerial Vehicles (UAV), as developed within the DLR ARTIS (Autonomous Research Testbed for Intelligent System) research fleet, especially those equipped with autonomous decision-making capabilities, are complex embedded systems with a multitude of sensors, significant modularization in system design, and have high safety requirements.

In this case study, we report on the application of stream-based monitoring with Lola to the DLR ARTIS UAV fleet. 

This includes a number of usability improvements to both the specification language in terms of syntax extensions as well as the user interface of the monitoring tool, to give better and more useful feedback at run-time.
We present representative specifications for different component modules of the system and report on our experiments integrating both offline and online monitoring into the system development process.
}

\section{Related Work}
In the area of unmanned aerial systems, earlier work on applying runtime verification has been performed by Schumann, Rozier et. al. at NASA \cite{DBLP:conf/rv/GeistRS14,DBLP:conf/tacas/ReinbacherRS14,DBLP:conf/rv/SchumannMR15}. The key differences to our approach are our use of a stream-based specification language with direct support for statistics, and that our framework uses a software-based instrumentation approach, which gives access to the internal state of system components. For non-assured control systems, a runtime supervision approach has been described in \cite{Gross2017}. A specific approach with a separate, verified hardware system to enforce geo-fencing has been described in \cite{Dill2016}.
Specification languages for runtime monitoring with similar expressivity include for example eSQL as implemented in the BeepBeep system \cite{DBLP:conf/bigdataconf/HalleGK16} and the Copilot language in \cite{DBLP:conf/rv/PikeNW11}, which has been applied to monitor airspeed sensor data agreement.

\section{Stream Runtime Monitoring}

\lola~is a stream-based specification language for monitoring, first presented for monitoring synchronous circuits in \cite{Angelo+others/05/Lola}, but more recently also used in the context of network monitoring \cite{DBLP:conf/rv/FaymonvilleFST16}.  
\lola~is a declarative specification mechanism based on stream equations and allows the specification of correctness properties as well as statistical properties of the system under observation. 

A \lola~specification consists of a set of stream equations, which define a set of \emph{input} streams, i.e. the signals of the system to the monitor, and a set of \emph{output} streams, whose values are defined by stream expressions and have access to past, present, and future values of the input streams and other output streams. All streams have a synchronous clock and evolve in a uniform way.

Since the language includes streams with numeric types and the stream expressions allow arithmetic expressions, it is easy to specify incrementally computable statistics.
Algorithms for both offline and online monitoring of \lola~specifications exist. In online monitoring, future values in stream expressions are evaluated by delaying the evaluation of their output streams.

Consider the following example specification.
\vspace{-2mm}
\begin{verbatim}	
input  bool valid
input  double height
output double m_height 
         		:= if valid { max(m_height[-1,0.0],height) } 
            	   else  { m_height[-1,0.0] }
\end{verbatim}
Here, given the input streams \textit{valid} and \textit{height}, the maximum valid height \textit{m\_height} is computed by taking the maximum over the previous \textit{m\_height} and the current \textit{height} in case the height is \textit{valid}, otherwise the previous value of \textit{m\_height} is used.
In \lola, the offset operator $s[x,y]$ handles the access to previous ($x<0$), present ($x=0$), or future ($x>0$) stream values of the stream $s$. 
The default value $y$ is used in case an offset $x$ tries to access a stream position past the end or before the beginning of a stream.
 
In this section, we present syntactic \lola~extensions, which were introduced to adapt the language to the domain-specific needs of monitoring unmanned aerial vehicles.
For formal definitions of the syntax and semantics of the base language syntax and semantics, we refer to \cite{Angelo+others/05/Lola} due to space reasons and will restrict ourselves to the extensions. 

\paragraph{Extensions to Lola}

A \lola~Specification is a system of equations of stream expressions over typed stream variables of the following form:
\vspace{-2mm}
\begin{align*}	
    \textbf{input}&\ T_1\ t_1\\ \vspace{-5mm}
    \dots   \\\vspace{-5mm}
    \textbf{input}&\ T_m\ t_m\\\vspace{-5mm}
	\textbf{output}&\ T_{m+1}\ s_1   :=  e_1(t_1,\dots,t_m, s_1, \dots s_n)&  \\ \vspace{-5mm}
	\dots   \\\vspace{-5mm}
	\textbf{output}&\ T_{m+n}\ s_n   :=  e_n(t_1,\dots,t_m, s_1, \dots s_n) &
\end{align*} 
The \emph{independent} stream variables $t_1,\dots,t_m$ with types $T_1,\dots,T_m$ refer to input streams, and the \emph{dependent} stream variables $s_1,\dots,s_n$ with types $T_{m+1},\dots,T_{m+n}$ refer to output streams. The \emph{stream expressions} $e_1,\dots,e_n$ have access to both input streams and output streams. To construct a stream expression $e$, Lola allows constants, functions, conditionals, and offset expressions to access the values of other streams.
Additionally, Lola specifications allow the definition of \emph{triggers}, which are conditional expressions over the stream variables. They generate notifications whenever they are evaluated to \emph{true}.

For a given valuation of input streams $\tau = \langle\tau_1,\dots,\tau_m\rangle$ of length $N+1$, the evaluation over the trace is defined as a stream of $N+1$ tuples $\sigma = \langle\sigma_1,\dots,\sigma_n\rangle$ for each dependent stream variable $s_i$, such that for all $1 \leq i \leq n$ and $0 \leq j \leq N$, if the equation $\sigma_i(j) = val(e_i)(j)$ holds, then we call $\sigma$ an \emph{evaluation model} of the Lola specification for $\tau$. The definition of $val(e_i)(j)$ is given in \cite{Angelo+others/05/Lola} as a set of partial evaluation rules, which tries to resolve the stream expressions as much as possible for each incoming event.

For the two extensions described here, we extend the definition of the stream expressions $e(t_1,\dots,t_m, s_1, \dots s_n)$ and the function $val$ as follows :
\begin{itemize}
\item Let $a$ be keyword of type $T$ (e.g.~\texttt{position}, \texttt{int\_min}, \texttt{int\_max} representing the maximal representable numbers), then $e = a$ is an atomic stream expression of type $T$.
    This adds direct access to system dependent maximal values for \texttt{int\_min, int\_max, double\_min, double\_max}, which is useful for default values. Additionally, we add direct access to the current stream position via $val(\texttt{position})(j) = j$.
\item Let $e'$ be a stream expression of type $T$, $d$ a constant of type $T$, and $i$ a positive int, then $e = e'\#[i,d]$ is a stream expression of type $T$. The value of this absolute offset is defined as,\\
$val(e\#[p,d])(j) =
	\begin{cases}
		val(e)(p) &\text{if}\ 0 \leq p \leq N\\
		val(d)(j) &\text{otherwise}
	\end{cases}$\\
The \emph{absolute offset operator} $\#$ refers to a position in the trace not relative to the current position, but instead absolute to the start of the trace.
\end{itemize} 
Common abbreviations:
\vspace{-3mm}
\begin{itemize}
\item[$\bullet$]\texttt{const T s := a $\widehat{=}$ output T s := a}
\item[$\bullet$]\texttt{ite($e_1$,$e_2$,$e_3$) $\widehat{=}$ if $e_1$\{$e_a$\}else\{$e_b$\}}
\item[$\bullet$]\texttt{if $e_1$\{$e_a$\} elif $e_2$\{$e_b$\} else\{$e_c$\} $\widehat{=}$ if $e_1$\{$e_a$\} else\{if $e_2$\{$e_b$\} else\{$e_c$\}\}}
\item[$\bullet$]\texttt{if $e_a=c_1$\{$e_1$\} elif $e_a=c_2$\{$e_2$\}$\dots$elif $e_a=c_n$\{$e_n$\} else\{$e_d$\} $\widehat{=}$ \newline switch $e_a$\{ case $c_1$\{$e_1$\} case $c_2$\{$e_2$\} $\dots$ case $c_n$\{$e_n$\} default\{$e_d$\}\}}
\end{itemize}

We have also added an extended switch operator, where the switch conditions have to be monotonically ordered. The semantics for this extended switch condition allows us to short-circuit the evaluation of large case switches.
There, the evaluation of \emph{lower} cases is omitted which helps e.g.~for properties on different flight phases with a large case split (take off, flight, landing, \dots) often encountered in the encoding of state machines. 

\paragraph{Usability Extensions}

We differentiate between two kinds of user feedback. On the one hand, we have online feedback, where notifications are displayed during the execution of the monitoring tool, on the other hand offline feedback, which creates another log-file for further post-analysis. This log-file can then in return be processed individually by the monitoring tool again, and is useful to first extract sections of interest and then process them later in detail.\\ 
\emph{Online Feedback} - Syntax:\qquad \texttt{obs\_kind condition message}\\
where \texttt{condition} is a boolean expression, and \texttt{message} is an arbitrary string.\\
The semantics for the different \texttt{obs\_kind}  is defined as follows:
\vspace{-2mm}
\begin{itemize}
\item \texttt{trigger}: Prints the \texttt{message} whenever the \texttt{condition} holds.
\item \texttt{trigger\_once}: Prints the \texttt{message} only the first time the \texttt{condition} holds.
\item \texttt{trigger\_change}: Prints the \texttt{message} whenever the \texttt{condition} value changes.
\item \texttt{snapshot}: Prints the monitor state, i.e.~the current stream values. 
\end{itemize}
\emph{Offline Feedback} - Syntax: \texttt{tag as} $y_1,\dots,y_n$ \texttt{if} $cond$ \texttt{with} $x_1,\dots,x_n$ \texttt{at} $l$\\
where $x_1,\dots,x_n$ are stream variables, $y_1,\dots,y_n$ are pairwise distinct stream names for the new log-file, and $cond$ is a boolean expression.
The semantics are as following: Whenever $cond$ holds, the value of $x_i$ is written to the respective $y_i$ column in the new log-file at location $l$.
These operations are especially interesting in offline post-flight analysis where they can ease the reasoning by generating enhanced log-files or by filtering the bulk of data to relevant fragments.

A special variant of this \emph{tagging} operator is \emph{filtering}, defined syntactically as:\\
\texttt{filter } $s_{1},...,s_{n}$ \texttt{ if } $cond$~~~~~~~~~~~~~~~~~~~~~~~\texttt{ at } $l$ ~~~~~:=\\
\texttt{tag as } $s_{1},...,s_{n}$ \texttt{ if } $cond$ \texttt{ with } $s_{1},...,s_{n}$ \texttt{ at } $l$ 
\\This operator copies all input streams to a new log-file, but filters on $cond$.

The syntax of \lola~permits not well-defined specifications, i.e.~where no unique evaluation model exists. Since the requirement of a unique evaluation model is a semantic criterion and expensive to check, we check a stronger syntactic criterion, namely \emph{well-formedness}, which implies well-definedness.
The well-formedness check can be performed on the \emph{dependency graph} for a \lola~specification, where the vertices of the multi-graph are stream names, and edges represent accesses to other streams. Weights on the edges are defined by the offset values of the accesses to other streams. As stated in \cite{Angelo+others/05/Lola}, if the dependency graph of a specification does not contain a zero-weight cycle, then the specification is well-formed.
If the dependency graph additionally has no positive-weight cycles, then the specification falls into the \emph{efficiently monitorable} fragment. Intuitively, this means it does not contain unbounded lookahead to future values of streams.


\paragraph{Implementation}

For the study, Lola has been implemented in C. Since most of the specifications discovered during this case study fall into the efficiently monitorable fragment of Lola, we focused on tuning the performance for this fragment.
In \cite{Angelo+others/05/Lola}, the evaluation algorithm maintains two equation stores.
Resolved equations are stored in $R$ and unresolved equations are stored in $U$.
Unresolved equations are simplified due to partial evaluation, rewrite, and substitution rules and, if resolved, added to $R$. Evaluated stream expressions are removed from $R$ whenever their largest offset has passed.
Our implementation uses an array for $R$ and an inverted index for $U$, for convenience we call the array for $R$ \emph{past array} and the inverted index for $U$ \emph{future index}.
By analyzing the dependency graph, we are able to calculate the size of the past index and can therefore pre-allocate a memory region \emph{once} on initialization.
The future index stores as keys the awaiting stream values and maps them to the respective waiting streams.
Here, we use the dependency graph to determine a fixed stream evaluation order to minimize the accesses to yet unresolved stream values.
Both data structures offer a fast access to values, the past index due to smart array indexing based on a flattening of the syntax tree of the stream expressions and the future index due to a simple lookup call for streams waiting on the resolution of a value.

%
%

\section{ARTIS Research Platform}
\begin{figure}[hb]
\begin{minipage}[t]{0.45\textwidth}
	\centering
	\includegraphics[width=.95\textwidth, clip]{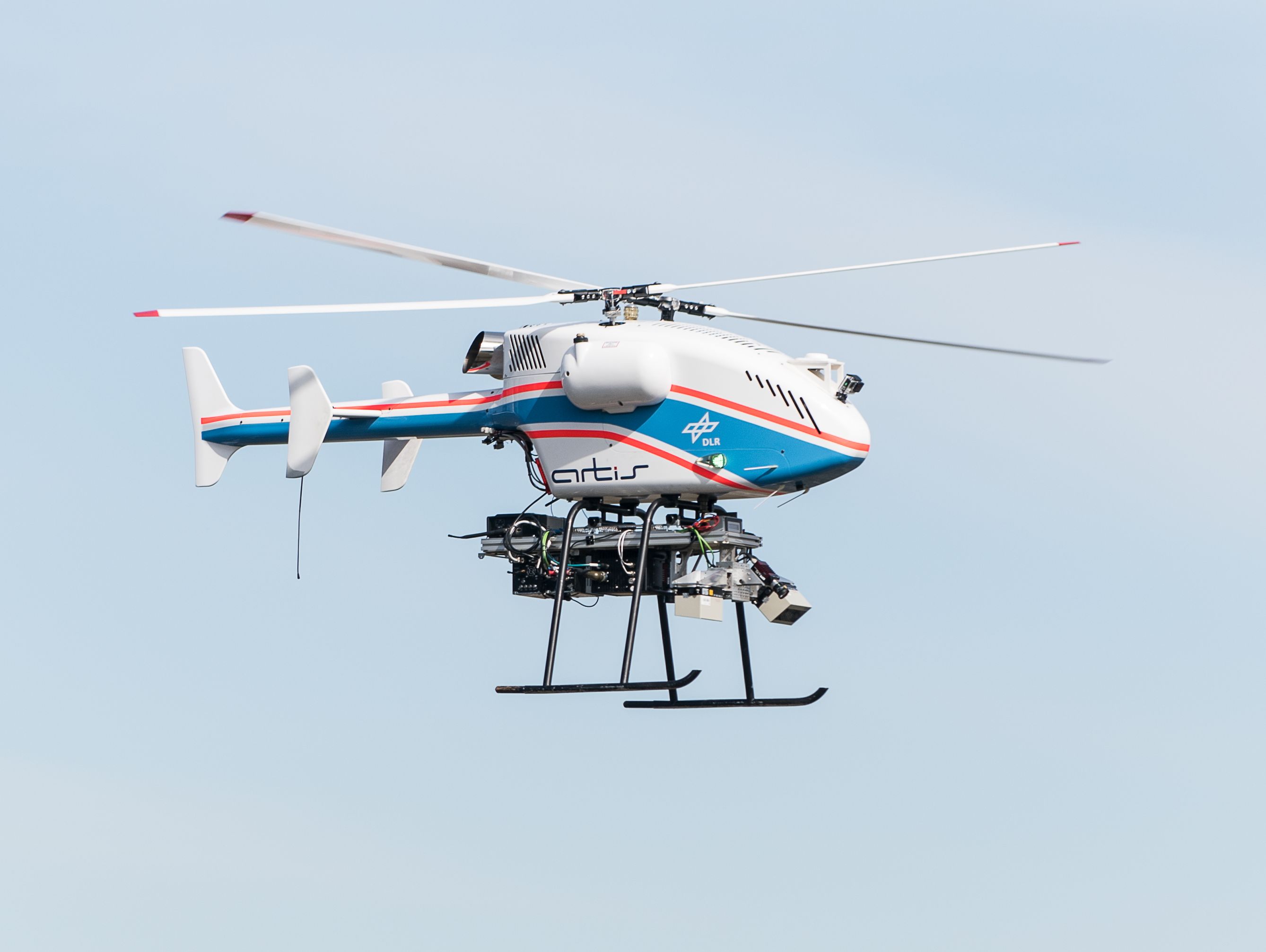}
 	\captionof{figure}{One example UAS of the DLR Unmanned Aircraft Fleet: SuperARTIS, a dual rotor configuration vehicle, shown with complete flight computers and sensor setup.}
 \label{f:super-artis}
\end{minipage}
\hfill
\begin{minipage}[t]{0.45\textwidth}
	\centering
	\includegraphics[width=.95\textwidth, clip]{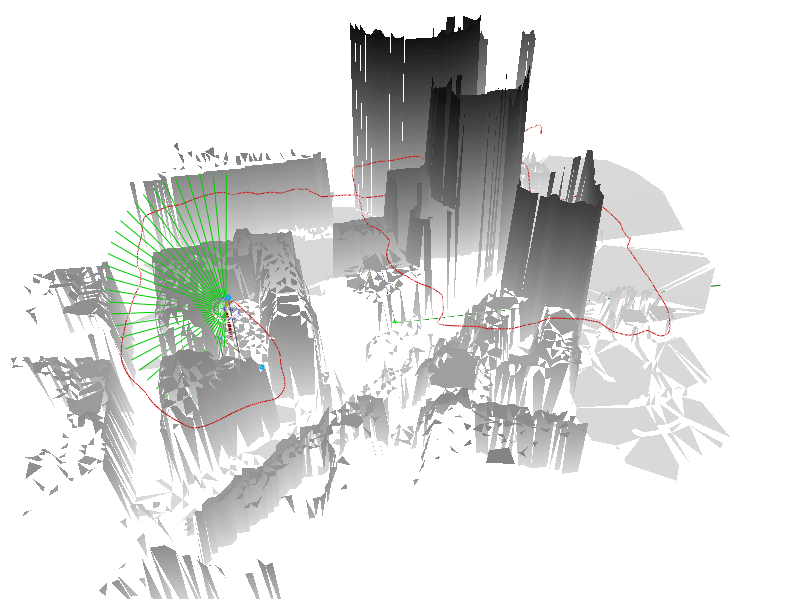}
 	\caption{A simulation of autonomous exploration of an unknown area with onboard perception: path flown (red), virtual distance sensor (green), obstacles mapped (grayscale).}
 \label{f:exploration}
\end{minipage}
\end{figure}

The DLR is researching UAS, especially regarding aspects of system autonomy and safety, utilizing its ARTIS platform.
The software framework enables development and test of intelligent functions that can be integrated and used with a whole fleet of UAS, comprised of several classes of aircraft with different technical equipment and autonomous capabilities. The latest addition to the DLR fleet is superARTIS, with a maximum take-off weight of 85 kg, Fig.~\ref{f:super-artis}.
A number of highly automated subsystems enables unmanned operations and provides required onboard functionality. 
As strict regulations apply, significant efforts in verification and validation activities are required to comply with standards to ensure functional correctness and safety. 
For the platform, there has been previous work on software verification as well as certification aspects for UAS\cite{Torens2013a,Torens2013b,Torens2014,Torens2015}.

Recently published supplements\cite{DO-333} to existing development standards for safety critical software\cite{DO-178C} introduced a regulatory framework to apply formal methods for the verification of aircraft.
However, due to a lack of expertise, there are some barriers for introduction of formal methods in industry\cite{Davis2013}.
Within our cooperation, starting with the use of runtime monitoring, the goal is to gradually introduce formal methods into the ARTIS development.
The use of runtime monitoring can support several aspects of research, development, verification, the operation, and even the autonomy of an unmanned aircraft.

\begin{figure}[hb]
\centering
\includegraphics[width=.9\textwidth]{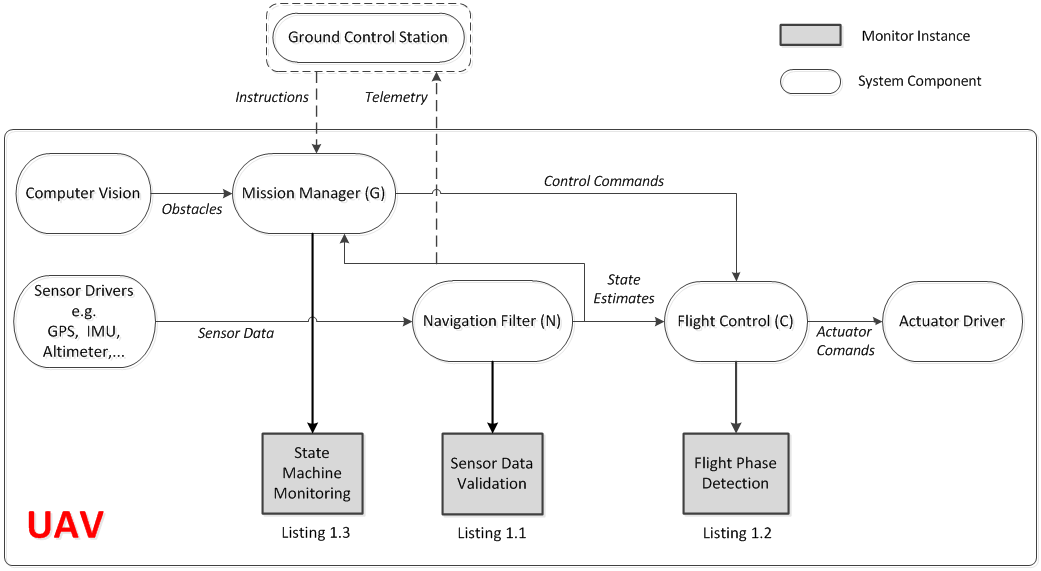}
\caption{ARTIS system overview. \vspace{-12pt}}
\label{fig:systemdiagram}
\end{figure}

ARTIS is based on a \textit{guidance, navigation, control (GNC)} architecture as illustrated in Fig.~\ref{fig:systemdiagram}, to be able to define high-level tasks and missions while also maintaining low-level control of the aircraft.

The \textit{flight control} realizes the control layer, that uses a model of the flight behavior to command the actuators so that the aircraft achieves desired position, height, speed, and orientation.
This system has to cope with external disturbances and keep the aircraft precisely on a given trajectory.
The \textit{navigation filter} is responsible for sensor input, such as GPS, inertial measurement, magnetic measurement.
The main task of the navigation filter module is the fusion of this heterogeneous information into consistent position information. 
The top-level component of the guidance layer is the \textit{mission manager}, which does mission planning and execution, breaking high-level objectives such as the exploration and mapping of an unknown area into actionable tasks by generation of suitable waypoints and the path planning to find an optimal route, Fig.~\ref{f:exploration}.

The three-tier architecture has the advantage of different abstraction layers that can be interfaced directly such that each layer represents a level of system autonomy.
The ARTIS unmanned aircraft have been evaluated in flight tests with respect to closed-loop motion planning in obstacle rich environments.

In flight, all modules are continuously generating an extensive amount of data, which is logged into files together with sensor data. This logging capability is important for the post flight traceability.
Log analysis can however quickly become infeasible due to interacting subsystems and possibly emergent aspects of the system. Data from log-files has to be processed for analysis and may need to be correlated with a context from a different module.
Runtime monitoring can support these aspects, by automating the analysis of log-files for specific properties.
An important feature is to filter only relevant log data, according to specific properties of the observation and tag it with a keyword for further analysis.
These properties can be introduced before conducting an experiment or simulation, or after the fact, since all relevant data is being saved to a log-file.


The software framework consists of a large code base with many modules and interfaces, which is under constant development. Interface changes triggered by a change in a single module are a significant problem during development, since other modules rely on implicit interface assumptions. Those assumptions together with a specification of a module can be monitored to detect inconsistencies early. They include internal properties of a subsystem, interface assumptions of a module, but also environmental assumptions of the overall aircraft. 

The explicit specification of assumptions is useful for system documentation, testing and verification, but also for the operation of the aircraft itself. In contrast to simple unit testing, a property that is formalized into a runtime monitor can not only be tested in a test fixture, but scale to the test phase.
Runtime monitoring can be integrated into software- and hardware-in-the-loop simulations as well as full-scale flight tests,  allowing the direct analysis for complex properties.
Since a formal specification mechanism is used, this also allows some reuse in other formal verification methodologies, e.g.~for LTL model checking.

System failures often occur not due to implementation errors, but because of the inherent difficulty of specifying all relevant system requirements correctly, completely, and free of contradictions. 
In nominal operation, the system behaves correctly, but as unforeseen events occur and environment assumptions are no longer valid, the system fails. 
Integrating runtime monitoring into normal system operation allows to continuously monitor for environmental assumptions and abnormal system behavior and allows to initiate contingency and failsafe procedures in case of unforeseen events.
In particular, EASA \cite{EASA2015a,EASA2015b} and JARUS\cite{JARUS2016} are working on concepts for the integration of unmanned aircraft into airspace that rely on the definition of specific operational limitations.
Here, the certification is no longer only dependent on the aircraft, but on the combination of it with the specific operation.
Limitations, such as constraining the altitude, the speed, the geographic location, etc.~of the operation can have a significant impact on the actual safety risk that emerges from the unmanned aircraft.
Runtime verification can support the safety of the aircraft by monitoring these limitations.



\section{Representative Specifications}
\label{sec:specs}
In this section, we will give some insights into representative classes of specifications, which have been used for stream runtime monitoring within the DLR ARTIS UAS fleet.
 These range from low-level sensor data validation and ensuring that the product of the sensor fusion is within the assumptions of the later stages of the control loop, to computing flight phase statistics, and validating the state machine of the mission planning and execution framework.
The specifications have been obtained by interviewing the responsible engineers for each component and collaborative specification writing. They have been validated by offline analysis of known test runs. The full set of specifications developed in the study can be found in \cite{SebastianSchirmer2016}.
Note that due to the time-triggered nature of the system design, all incoming signals to the monitor are timestamped and arrive with a frequency of 50 Hz. 
For the following specifications, the integration of the respective monitor instance is depicted in Figure \ref{fig:systemdiagram}.

\subsection{Sensor Data Validation}
\begin{listing}[H]
\begin{minted}[linenos,
  fontsize=\scriptsize]{lola}
input  double lat, lon, ug, vg, wg, time_s, time_micros
output double time := time_s + time_micros / 1000000.0 
output double flight_time := time - time#[0,0.0]
output double frequency := switch position{
                            case 0  { 1.0 / ( time[1,0.0] - time ) }
                            default { 1.0 / ( time - time[-1,0.0] ) } }
output double freq_sum := freq_sum[-1,0.0] + frequency
output double freq_avg := freq_sum / double(position+1)				 
output double freq_max := max( frequency, freq_max[-1,double_min] )
output double freq_min := min( frequency, freq_min[-1,double_max] )

output double velocity := sqrt( ug^2.0 + vg^2.0 + wg^2.0 )
const  double R        := 6373000.0
const  double pi       := 3.1415926535

output double lon1_rad := lon[-1,0.0] * pi / 180.0
output double lon2_rad := lon * pi / 180.0
output double lat1_rad := lat[-1,0.0] * pi / 180.0
output double lat2_rad := lat * pi / 180.0
	
output double dlon     := lon2_rad - lon1_rad
output double dlat     := lat2_rad - lat1_rad
output double a := (sin(dlat/2.0))^2.0  +  
                   cos(lat1_rad) * 
                   cos(lat2_rad) * 
                   (sin(dlon/2.0))^2.0
output double c := 2.0 * atan2( sqrt(a), sqrt(1.0-a) )
output double gps_distance := R * c

output double passed_time     := time - time[-1,0.0]
output double distance_max    := velocity * passed_time
output double dif_distance    := gps_distance - distance_max
const  double delta_distance  := 1.0
output bool   detected_jump   := switch position { 
                                  case 0 { false } 
                                  default { dif_distance >  delta_distance } }
snapshot detected_jump with "Invalid GPS signal received!"
\end{minted} 
 \caption{The specification used for Sensor Data Validation}
  \label{spec:sensordatavalidation}
\end{listing}
In Listing~\ref{spec:sensordatavalidation}, we validate the output of the navigation part of the sensor fusion.
Given the sensor data, e.g.~GPS position and the inertial measurement, the navigation filter outputs vehicle state estimates, depicted in Figure \ref{fig:systemdiagram}.
Based on this specification, the monitor checks the frequency of incoming signals and detects GPS signal jumps.
The detection of frequency deviations can point to problems in the signal-processing chain, e.g.~delayed, missing, or corrupted sensor values (Line 4 to 10).
Since the vehicle state estimation is used for control decisions, errors would propagate through the whole system.
From Line 21 to 37, the GPS signal jumps are computed by the Haversine formula.
It compares the traveled distance, first by integrating over the velocity values received by the IMU unit, and second by computing the distance as measured by the GPS coordinates. All calculations are performed in \lola~and compared against a threshold.
Since the formula expects the latitude and longitude in radians and we receive them in decimal degree, we convert them first (Line 16 to 19).

\subsection{Flight Phase Detection}
\begin{listing}[H]
\begin{minted}[linenos,
  fontsize=\scriptsize]{lola}
input  double time_s, time_micros, vel_x, vel_y, vel_z, 
           	 fuel, power, vel_r_x, vel_r_y, vel_r_z 
output double time := time_s + time_micros / 1000000.0
output double flight_time := time - time#[0,0.0]
output double frequency := switch position{
                            case 0  { 1.0 / ( time[1,0.0] - time ) }
                            default { 1.0 / ( time - time[-1,0.0] ) } }
output double freq_sum := freq_sum[-1,0.0] + frequency
output double freq_avg := freq_sum / double(position+1)				 
output double freq_max := max( frequency, freq_max[-1,double_min] )
output double freq_min := min( frequency, freq_min[-1,double_max] )

const  double vel_bound       := 1.0
output double velocity       := sqrt( vel_x^2.0 + vel_y^2.0 + vel_z^2.0 )
output double velocity_max   := if reset_max[-1,false] { velocity } 
                              else { max( velocity, velocity_max[-1,0.0]) }
output double velocity_min   := if reset_max[-1,false] { velocity } 
                              else { min( velocity, velocity_min[-1,0.0]) }	
output double dif_max        := difference(velocity_max, velocity_min) 
output bool   reset_max      := dif_max > vel_bound
output double reset_time     := if reset_max | position = 0 { time } 
                              else  { reset_time[-1,0.0] }
output int unchanged         := if reset_max[-1,false] { 0 } 
                              else { unchanged[-1,0] + 1 } 
snapshot unchanged = 150 with "Phase detected!"

output double  vel_dev := difference(vel_r_x,vel_x) + difference(vel_r_y,vel_y) 
                          + difference(vel_r_z,vel_z) 
output double  dev_sum   := vel_dev + dev_sum[-1,0] 
output double  vel_av    := dev_sum / double((position+1)*3)
output int worst_dev_pos := if worst_dev[-1,double_min] < vel_dev { position } 
                          else { worst_dev_pos[-1,0] }
output double worst_dev  := if worst_dev[-1,double_min] < vel_dev { vel_dev }  
                          else { worst_dev[-1,0.0] }

output double fuel_p  := ( ( fuel#[0,0.0] - fuel ) /  (fuel#[0,0.0]+0.01)  )
output double power_p := ( (power#[0,0.0] - power) /  (power#[0,0.0]+0.01) )
trigger_once fuel_p  < 0.50  with "Fuel below half capacity"
trigger_once fuel_p  < 0.25  with "Fuel below quarter capacity"
trigger_once fuel_p  < 0.10  with "Urgent: Refill Fuel!"
trigger_once power_p < 0.50  with "Power below half capacity"
trigger_once power_p < 0.25  with "Power below quarter capacity"
trigger_once power_p < 0.10  with "Urgent: Recharge Power!"
 \end{minted} 
 \caption{The specification used for Flight Phase Detection}
  \label{spec:flightphasedetection}
\end{listing}
The mission manager describes high-level behaviors, e.g.~the hover or the fly-to behavior.
The hover behavior implies that the aircraft remains at a fixed location whereas the fly-to behavior describes the movement from a source to a destination location.
These high-level behaviors are then automatically synthesized into low-level control commands which are send to the flight control.
Given the state estimation, the flight control then smoothes the control commands into applicable actuator commands, $($\emph{vel\_r\_x}, \emph{vel\_r\_y}, \emph{vel\_r\_z}$)$, depicted in Figure \ref{fig:systemdiagram}.
Hence, the actuator commands implement the desired high-level behavior.
Since the actuator movements are limited and therefore smoothed by the flight control, there is a gap between the control commands and the actuator commands. 

In Listing \ref{spec:flightphasedetection}, monitoring is used to recognize flight phases where the velocity of the aircraft stays below a small bound for longer than three seconds (Line 13 to 25).
In post-flight analysis, the recognized phases can be compared to the high-level commands to validate the suitability of the property for the respective behavior.
Furthermore, from Line 3 to 11, the frequency is examined and, from Line 27 to 34, the
deviations between the reference velocity, given by the flight controller, and the actual velocity $($\emph{vel\_x}, \emph{vel\_y}, \emph{vel\_z}$)$ are detected. 
Bound checks on fuel and power detect further boundary conditions and produce notifications for the difference urgency levels.

\subsection{Mission Planning and Execution}

\begin{listing}[H]
\begin{minted}[linenos,
  fontsize=\scriptsize]{lola}
input double time_s, time_micros
input int stateID_SC, OnGround
const int Start                 := 0
const int MissionControllerOff  := 1
                                   ...
const int HammerHeadTurn        := 16

output double time := time_s + time_micros / 1000000.0
output double flight_time := time - time#[0,0.0]

output bool change_state := switch position { 
                              case 0 { false } 
                              default { stateID_SC != stateID_SC[-1,-1] } }
trigger change_state

output string state_enum := switch stateID_SC  {
                              case 0 { "Start" }
                              case 1 { "MissionControllerOff" }
                               ...
                              case 16{ "HammerHeadTurn" }
                              default{ "Invalid" }  }
output string state_trace := 
   switch position {  case 0  { state_enum }  default { 
   if change_state { concat(concat(state_trace[-1,""]," -> "),state_enum) } 
   else { state_trace[-1,""] } }  }
   
output double entrance_time  := if change_state { time } 
                              else { entrance_time[-1,0.0] }
const double landing_timebnd := 20.0
output double landing_info   := if stateID_SC = Landing { 0.0 } 
                              else { time - entrance_time[-1,0.0] }
output bool landing_error    := stateID_SC = Landing & OnGround != 1  &  
                                landing_info > landing_timebound

 \end{minted}  
 \caption{The specification used for Mission Planning and Execution}
  \label{spec:miplex}
\end{listing}

The mission planning and execution component within the ARTIS control software is responsible for executing uploaded flight missions onboard the aircraft.
The received mission from the ground control station is planned based on the current state estimate of the vehicle and the sensed obstacles, depicted in Figure \ref{fig:systemdiagram}.
The mission manager and its internal planner essentially consists of a large state machine which controls the parameters of the current flight state. In the corresponding specification seen in Listing~\ref{spec:miplex}, the state machine is encoded into the specification, and the state trace is recovered from the inputs and converted to a readable form. Starting in Line 27 of the specifications, we record entrance and exit times of certain flight states, and check a landing time bound to ensure a bounded liveness property.
With this specification, we are able to detect invalid transition and ensure a landing time bound.
In an extended version, we further specified properties for each location.
Specifically, we aggregated statistics on the fuel consumption, the average velocity, and the maximal, average, and total time spend in the respective location.

\section{Monitoring Experiments}

\subsection{Offline Experiments}
The experiments indicate how \lola~performs with the given set of specifications in an offline monitoring context. 
This mode was especially useful during specification development, because a large number of flight log-files was readily available and could be used to debug the developed specifications.

\begin{figure}
\vspace{-12pt}
\centering
\includegraphics[scale=0.4]{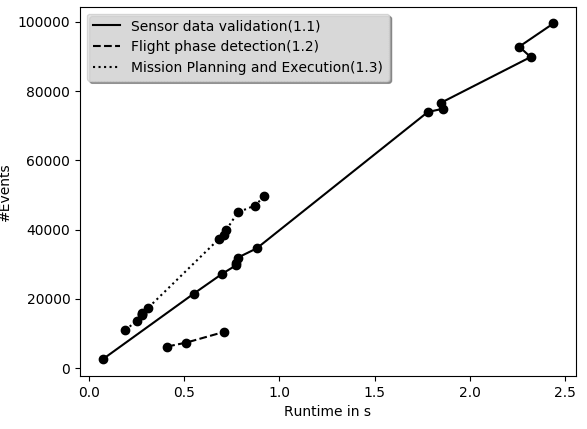}
\caption{The results of the offline experiments for the specifications presented in Section~\ref{sec:specs}.\vspace{-12pt}}
\label{fig:offlineResults}
\end{figure}

As offline parameters, \lola~receives one or more specifications and the log data file.
As an optional parameter, a location for the monitor output can be set.
The offline experiments were conducted on a dual-core machine with an 2.6GHz Intel Core i5 processor with 8GB RAM.  The input streams files were stored on an internal SSD.
Runtime results are shown in Figure~\ref{fig:offlineResults}. Memory consumption was below 1.5 MB. The simulated flight times range up to 15 minutes.

The experimental results show that our implementation could process the existing log-files in this application context within seconds. Using an additional helper tool, which automatically runs offline monitoring on a set of given log-files, and further specifications, we were able to identify system boundaries and thresholds without further effort. 

\subsection{Online Experiments} \label{sec:Online}

The \lola~online interface is written in C++.
Due the absence of both a software bus and a hardware bus, the monitor interface is coupled to the existing logging interface.
Therefore, we can use the created log-files to evaluate the monitor impact on the system by comparing the logged frequencies.

\begin{figure}[htb]
\adjustbox{valign=t}{\begin{minipage}[t]{0.48\textwidth}
	\includegraphics[width=.8\textwidth, clip]{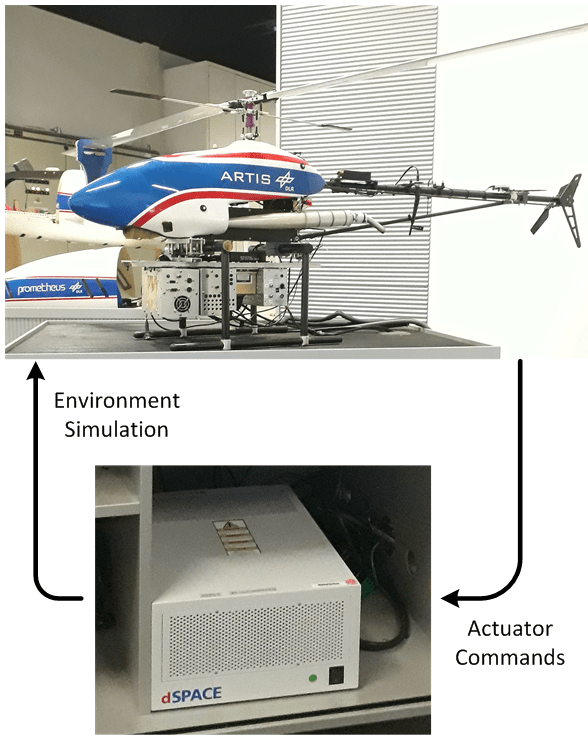}
	\label{fig:hitl}
	\caption{Test rig for hardware-in-the-loop (HITL) experiments.}
\end{minipage}}
\hfill
\adjustbox{valign=t}{\begin{minipage}[t]{0.48\textwidth}
	\includegraphics[scale=0.44]{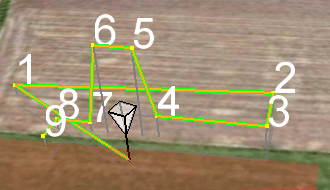}
	\includegraphics[scale=0.335]{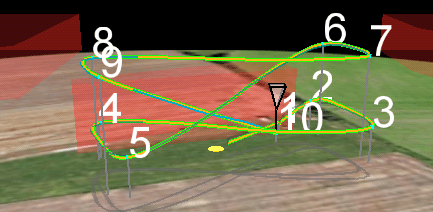}
	\caption{Tracked flight-paths of two scenarios. On top, we see a hover simulation and at the bottom a fly-to simulation.} 
\label{fig:simulationsHilandSil}
\end{minipage}}
\end{figure}

The Hardware-in-the-loop (HiL) experiments were run on a flight computer with an Intel Pentium with a 1.8GHz and 1GB RAM, running a Unix-based RTOS.
As inter-thread communication, we use shared memory for both the stream value delivery and the monitor output.
A simulated world environment and a flight mission were set, 
the worlds and the missions are depicted in Figure~\ref{fig:simulationsHilandSil}.
In the experiments, we monitored the system with a superset of the specifications described in Section~\ref{sec:specs}. 
We evaluated the impact of online monitoring as following.
For each mission, all experiments flew the same planned route without noticeable deviations, analyzed per manual inspection of an online visualization of the flight as seen in Figure~\ref{fig:simulationsHilandSil}.
To measure the system performance impact, we compare the average frequency determined by the timestamps in the monitor, if available for the component, otherwise with the frequency computed afterwards by offline monitoring on the logs of the experiment with the average frequency of a non-monitored execution.

The results for hardware-in-the-loop testing are given in Figure~\ref{hitlresults}.
\begin{figure}[htb]
\centering

	\resizebox{.65\linewidth}{!}{
\begin{tabular}{|r|c|c|c|c|c|c|c|c|}\hline
\textbf{monitor} (evalstep) & \multicolumn{3}{|c|}{\textbf{online} - AvgFreq} & \multicolumn{4}{|c|}{\textbf{offline} - AvgFreq} \\
\cline{2-8}
& nav & ctrl & mgr & gps-p & gps-v & imu  & mgn\\
& (Hz) & (Hz) & (Hz) & (Hz)  & (Hz) & (Hz) & (Hz) \\
\hline  \hline
\textit{No Monitor} & \textit{50.0} & \textit{50.0} & \textit{50.0} & 20.0 & 20.0 & 100.0 & 10.0  \\
\hline
nav\_monitor (1) & \textbf{50.0} & \textit{50.0} & \textit{50.0} & 20.0 & 20.0 & 100.0 & 10.0  \\
\hline
nav\_monitor (1) & \textbf{50.0} & - & \textit{50.0} & 20.0 & 20.0 & 100.0 & 10.0  \\
ctrl\_monitor (1) & - & \textbf{50.0} & - & " & " & " & "  \\
\hline
nav\_monitor (1) & \textbf{50.0} & - & - & 20.0 & 20.0 & 100.0 & 10.0  \\
ctrl\_monitor (1) & - & \textbf{50.0} & - & " & " & " & "  \\
mgr\_monitor (1) & - & - & \textbf{50.0} & " & " & " & "  \\
\hline
nav\_monitor (100) & \textbf{50.1} & - & - & 20.0 & 20.0 & 100.1 & 10.0  \\
ctrl\_monitor (100) & - & \textbf{50.3} & - & " & " & " &  " \\
mgr\_monitor (100) & - & - & \textbf{50.2} & " & " & " &  " \\
\hline
\end{tabular}
}
\caption{Monitor performance results. The first line denotes the reference values without monitors, and more monitors are added in the further lines. The evalstep parameter represents the amount of buffering of input values before evaluation is triggered. Bold values denote frequencies measured online with Lola, the other frequency values have been determined after the test by offline analysis.\vspace{-12pt}}
\label{hitlresults}
\end{figure}
The experimental results show that the timing behavior, i.e.~the frequencies, are minimally affected.
Thus, the current implementation is sufficiently fast for online monitoring in this experimental setting.
The monitoring approach can run aside the logging.
By setting the evalstep parameter to 100 to simulate a data burst for a monitor evaluation step, the sensitive time-triggered system is slightly affected.

\section{Conclusion}

We have presented our experience from a successful case study for stream-based runtime monitoring with Lola in a safety-critical application environment. The DLR ARTIS family of unmanned aerial vehicles provides a unique research testbed to explore the applicability of runtime monitoring in this application.

Our experiences show that the integration of runtime monitoring into an existing system has benefits for both the system and its development process. While the primary goal is to ensure the correct behavior of the system, monitor the health of the system components, and possibly trigger fail-safe strategies, there are a number of secondary effects on the system development process, which may aid the adoption of runtime verification techniques. Those benefits include time savings during system debugging, a common and faster way to perform log-file analysis, and a better documentation of interface assumptions of components. The re-use of the same specifications in all contexts increases their appeal.
The specification development has already influenced the language and system implementation of our stream-based specification language. The presented extensions improve the readability of specifications, and the insight that efficiently monitorable specifications suffice for online monitoring guides optimization efforts in the implementation. The equation-based, declarative specification style of Lola with a rich type and function supports its use in a real engineering project.

As the adaptation of autonomy into the system designs of regulated industries increases, we expect runtime monitoring to play an important role in the certification process of the future. Runtime verification techniques allow the deployment of trusted, verified components into systems in unreliable environments and may be used to trigger pre-planned contingency measures to robustly control hazardous situations. To perform these important tasks, the monitor needs to comprehensively supervise the internal state of the system components to increase the self-awareness into the system health and to trigger a timely reaction.

\bibliography{main}

	
\end{document}